\documentclass{article}
\usepackage{amssymb}

\begin{document}
\title{Geodesic Completeness of Orthogonally Transitive Cylindrical Spacetimes}
\author{L. Fern\'andez-Jambrina\\
Departamento de Ense\~nanzas B\'asicas de la Ingenier\'{\i}a Naval\\
E.T.S.I. Navales\\ Arco de la Victoria s/n\\ E-28040-Madrid }
\date{}

\maketitle
\begin{center}PACS numbers: 04.20 Dw
\end{center}
\begin{abstract}
 In this paper a theorem is derived in order to provide  a wide sufficient
condition for an orthogonally transitive cylindrical spacetime
to be singularity-free. The applicability of the theorem is tested on 
examples provided by the literature that are known to have regular curvature
invariants.
\end{abstract}
\newpage

\section{Introduction}

The issue of establishing whether a Lorentzian manifold is geodesically
complete is not in principle a simple one since there is no Hopf-Rinow theorem
that could settle the matter as it happens in the Riemannian case. One could
think that regularity of the curvature invariants might be helpful, but there
are known examples of spacetimes with regular invariants, such as Taub-NUT,
\cite{HE}, that  enclose geodesics that are not complete in their affine
parametrization due to a phenomenon called imprisoned incompleteness.

Taking completeness of causal geodesics ($g$-completeness) as a definition of
absence of singularities (no observer in free fall leaves the spacetime in a
finite proper time), one can resort to many theorems in the literature (cfr.
\cite{HE}, \cite{Beem} and references therein) in order to determine whether a
spacetime is singular. But on the contrary theorems that provide large families 
of nonsingular spacetimes are not very usual \cite{miguel}, \cite{Manolo} and
in principle the proof of geodesic completeness involves cumbersome
calculations \cite{Chinea}. 

Instead of dealing with general Lorentzian manifolds, we shall approach
orthogonally transitive cylindrical spacetimes \cite{class} since they have
provided many examples of regular manifolds (cfr. \cite{Esc}, \cite{grg}) in
inhomogeneous cosmology \cite{kra}. 

Our aim will be the generalization of the theorem on diagonal
orthogonally transitive cylindrical spacetimes in
\cite{Manolo} to nondiagonal models and thereby comprise all known nonsingular
cylindrical perfect fluid spacetimes in the literature. 

First we shall show that the second-order system of geodesic equations can be
reduced by the use of constants of motion to three first-order equations plus
two quadratures. This fact will simplify the analysis of the prolongability of
the geodesics and will enable us to write a sufficient condition for
completeness of orthogonally transitive cylindrical spacetimes in a theorem.

\section{Geodesic Equations}

We shall write the metric of an orthogonally transitive cylindrical spacetime in
a chart  using isotropic coordinates $t,r$ for the susbspace orthogonal to the
orbits of the isometry group  and coordinates $\phi$, $z$ adapted to the
commuting generators of the group of isometries. The metric shall be determined
by four functions, $g,f,A,\rho$, of the coordinates $t,r$,
\begin{eqnarray}
&ds^2=e^{2\,g(t,r)}\left\{-dt^2+dr^2\right\}+\rho^2(t,r)e^{2\,f(t,r)}d\phi^2
+\nonumber\\&+e^{-2\,f(t,r)}\{dz+A(r,t)\,d\phi\}^2.\label{metric}
\end{eqnarray}
and we shall assume that these functions are $C^2$ in their
range,

\begin{eqnarray}
-\infty<t,z<\infty,\quad 0<r<\infty,\quad 0<\phi<2\,\pi.
\end{eqnarray}

The axis will be located where the norm of the axial Killing field vanishes, 
\begin{equation}
0=\Delta=g(\xi,\xi)=\rho^2(t,r)e^{2\,f(t,r)}+e^{-2\,f(t,r)}\,A^2(r,t),
\end{equation}
which means that both $A$ and $\rho$ must vanish on the axis, since $f$ is a
smooth function.

Since the choice of isotropic coordinates is not unique, we can take
advantage of this freedom to have $r=0$ as the equation for the axis. In order
to avoid conical singularities, the usual requirement \cite{Kramer} will be
imposed in order to have a well defined axis,

\begin{equation}
\lim_{r\to
0}\frac{g({\rm grad}\,\Delta,{\rm grad}\,\Delta)}{4\,\Delta}=
1.\label{reg} \end{equation}

Denoting by a dot differentiation with respect to the affine parameter, 
two of the four second-order geodesic equations,
\begin{equation} 
\ddot x^i+\Gamma^i_{jk}\dot x^j\dot x^k = 0,\end{equation}
can be integrated taking into account that there are two first integrals  of the geodesic
motion associated with the generators of the isometries. These are the angular
momentum around the axis, $L$, and the linear momentum along the axis, $P$, of a test particle
of unit mass,
\begin{eqnarray}
L=e^{2\,f(t,r)}\,\rho^{2}(t,r)\,\dot\phi+
e^{-2\,f(t,r)}\,A(t,r)\,\{\dot z+A(t,r)\,\dot \phi\},
\end{eqnarray}

\begin{eqnarray}
P=e^{-2\,f(t,r)}\{\dot z+A(t,r)\,\dot \phi\}.
\end{eqnarray}

The affine parametrization is determined, up to an affinity of the real line, by the
prescription,
\begin{eqnarray}
\delta=e^{2\,g(t,r)}\left\{\dot
t^2-\dot  r^2\right\}-
\{L-P\,A(t,r)\}^2\rho^{-2}(t,r)e^{-2\,f(t,r)}-P^2e^{2\,f(t,r)},\label{delta}
\end{eqnarray}
where $\delta$ is one for timelike, zero for null and minus one for
spacelike geodesics. Since we are dealing just with causal geodesics, for
our purposes
$\delta$ will always be positive. After writing $\dot z$, $\dot \phi$ as functions of $L$ and
$P$, the second-order equations in $t$ and $r$, 
\begin{eqnarray}
\ddot t+g_t(t,r)\dot t^{2}
+2\,g_r(t,r){
\dot t}{\dot r}+g_t(t,r){\dot r}
^{2}-{P}^{2}{e^{2\,\left\{f(t,r)-g(t,r)\right\}}}f_t(t,r)
+\nonumber\\+e^{-2\,\left\{f(t,r)+g(t,r)\right\}}
\frac{\{L-PA(t,r)\}^{2}}
{\rho^2(t,r)}\left\{\frac{\rho_t(t,r)}{\rho(t,r)}+f_t(t,r)+
\frac{PA_t(t,r)}{L-PA(t,r)}\right\},
\label{eq1}
\end{eqnarray}

\begin{eqnarray}
\ddot r+g_r(t,r)\dot t^{2}
+2\,g_t(t,r){
\dot t}{\dot r}+g_r(t,r){\dot r}
^{2}+{P}^{2}{e^{2\,\left\{f(t,r)-g(t,r)\right\}}}f_r(t,r)
-\nonumber\\-e^{-2\,\left\{f(t,r)+g(t,r)\right\}}
\frac{\{L-PA(t,r)\}^{2}}
{\rho^2(t,r)}\left\{\frac{\rho_r(t,r)}{\rho(t,r)}+f_r(t,r)+
\frac{PA_r(t,r)}{L-PA(t,r)}\right\},\label{eq2}
\end{eqnarray}
can be  rearranged in a form that will be useful afterwards,

\begin{eqnarray}
\{e^{2\,g(t,r)}\dot t\}^{\cdot} -
e^{-2g(t,r)}F(t,r)F_t(t,r)=0,
\label{eeq1}
\end{eqnarray}

\begin{eqnarray}
\{e^{2\,g(t,r)}\dot
r\}^{\cdot}+e^{-2g(t,r)}F(t,r)F_r(t,r)=0,\label{eeq2}
\end{eqnarray}

\begin{equation}
F(t,r)=e^{g(t,r)}\sqrt{\delta+P^2e^{2f(t,r)}+
\{L-PA(t,r)\}^2\frac{e^{-2f(t,r)}}{\rho^{2}(t,r)}},
\end{equation}
that have the same estructure as in the diagonal case.

If at least one of the constants $L$, $P$, $\delta$ is different from zero, the system is
equivalent to three equations of first order for future-pointing (past-pointing) geodesics,
\begin{eqnarray}
\dot
t=\pm e^{-2g(t,r)}F(t,r)\cosh\xi(t,r),
\end{eqnarray}

\begin{eqnarray}
\dot
r=e^{-2g(t,r)}F(t,r)\sinh\xi(t,r),
\end{eqnarray}

\begin{eqnarray}
\dot\xi(t,r)=-
e^{-2g(t,r)}\left\{\pm F_t(t,r)\sinh\xi(t,r)+F_r(t,r)\cosh\xi(t,r)\right\},
\end{eqnarray}
parametrizing (\ref{delta}) by a function $\xi(t,r)$. More explicitly, the last equation takes
the form,
\begin{eqnarray}
&\dot\xi=-\displaystyle\frac{e^{-g}}{\sqrt{\delta+\Lambda^2\rho^{-2}e^{-2f}+P^2e^{2f}}}
\left\{\cosh\xi\left(\delta
g_r+\Lambda^2\frac{e^{-2f}}{\rho^2}h_r+
P^2e^{2f}q_r\right)\pm
\right.\nonumber\\&\pm\left.\sinh\xi\left(\delta
g_t+\Lambda^2\frac{e^{-2f}}{\rho^2}h_t+
P^2e^{2f}q_t\right)\right\},\label{xi}
\end{eqnarray}
\begin{equation}h=g-f-\ln{\rho}+\ln|\Lambda|,\quad \Lambda=L-PA,\quad q=g+f,\end{equation}
that will be useful for deriving prolongability conditions for causal geodesics. The minus
(plus) sign corresponds to past-pointing (future-pointing) geodesics. 

Note that the general
equations are obtained from those of the diagonal case just replacing $L$ by
$\Lambda$, that can be therefore considered as a sort of `effective angular momentum' in the
case where the Killing fields are not orthogonal.

\section{Prolongability of the geodesics}

In this section we shall introduce two theorems on causal geodesic completeness
of orthogonally transitive cylindrical spacetimes. Null coordinates,
\begin{equation}
u=\frac{t+r}{2},\qquad v=\frac{t-r}{2},\end{equation} will play an important role in the
results.

\begin{description}
\item[Theorem 1:] An orthogonally transitive cylindrical spacetime endowed with a metric whose
local expression in terms of $C^2$ metric functions $f,g,A,\rho$ is given by (\ref{metric})
such that the axis is located at $r=0$ has complete future causal geodesics if the following set
of conditions is fulfilled:

\begin{enumerate}
\item For large values of $t$ and increasing $r$, 
\begin{enumerate}
\item \label{Mxi1}$\left\{
\begin{array}{l}g_u\ge 0\\
h_u\ge 0\\
q_u\ge 0,\end{array}\right.$
\item \label{Mxi2} Either $\left\{
\begin{array}{lcl}{g_r}\ge 0&\textrm{or}& |g_r|\lesssim g_u\\h_r
\ge 0&\textrm{or}& |h_r|\lesssim h_u\\q_r\ge 0&\textrm{or}& |q_r|\lesssim
q_u.\end{array}\right.$
\end{enumerate}
\item For $L\neq0$ and large values of $t$ and decreasing $r$, 
\begin{enumerate}
\item \label{mxi1}$\delta\,g_v+P^2e^{2f}\,
q_v+\Lambda^2\frac{e^{-2f}}{\rho^2}h_v\ge 0$
\item \label{mxi2} Either $\delta g_r+P^2e^{2f}\,
q_r+\Lambda^2\frac{e^{-2f}}{\rho^2}\,h_r\le 0$ {or} $\delta g_r+P^2e^{2f}\,
q_r+\Lambda^2\frac{e^{-2f}}{\rho^2}\,h_r\lesssim \delta g_v+P^2e^{2f}\,
q_v+\Lambda^2\frac{e^{-2f}}{\rho^2}\,h_v.$
\end{enumerate}
\item \label{tt} For large values of the time coordinate  $t$, constants $a,b$ exist such that 
$\left.\begin{array}{c}2\,g(t,r)\\g(t,r)+f(t,r)+\ln\rho-\ln|\Lambda|\\
g(t,r)-f(t,r)\end{array}\right\}\ge-\ln|t+a|+b.$

\item \label{ax} The limit $\displaystyle\lim_{r\to 0}\frac{A}{\rho}$ exists.
\end{enumerate}

\end{description}

A theorem can be obtained for past-pointing geodesics just changing the sign of
the time derivatives in the  previous one.

\begin{description}
\item[Theorem 2:] An orthogonally transitive cylindrical spacetime endowed with
a metric whose local expression in terms of $C^2$ metric functions $f,g,A,\rho$
is given by (\ref{metric}) such that the axis is located at $r=0$ has complete
past causal geodesics if the following set of conditions is fulfilled:

\begin{enumerate}
\item For small values of $t$ and increasing $r$, 
\begin{enumerate}
\item $\left\{
\begin{array}{l}g_v\le 0\\
h_v\le 0\\
q_v\le 0,\end{array}\right.$
\item  Either $\left\{
\begin{array}{lcl}{g_r}\ge 0&\textrm{or}& |g_r|\lesssim -g_v\\h_r
\ge 0&\textrm{or}& |h_r|\lesssim -h_v\\q_r\ge 0&\textrm{or}& |q_r|\lesssim
-q_v.\end{array}\right.$
\end{enumerate}
\item For $L\neq0$ and small values of $t$ and decreasing $r$, 
\begin{enumerate}
\item $\delta\,g_u+P^2e^{2f}\,
q_u+\Lambda^2\frac{e^{-2f}}{\rho^2}h_u\le 0$
\item Either $\delta g_r+P^2e^{2f}\,
q_r+\Lambda^2\frac{e^{-2f}}{\rho^2}\,h_r\le 0$ {or} $\delta g_r+P^2e^{2f}\,
q_r+\Lambda^2\frac{e^{-2f}}{\rho^2}\,h_r\lesssim |\delta g_u+P^2e^{2f}\,
q_u+\Lambda^2\frac{e^{-2f}}{\rho^2}\,h_u|.$
\end{enumerate}
\item For small values of the time coordinate  $t$, constants $a,b$ exist such that 
$\left.\begin{array}{c}2\,g(t,r)\\g(t,r)+f(t,r)+\ln\rho-\ln|\Lambda|\\
g(t,r)-f(t,r)\end{array}\right\}\ge-\ln|t+a|+b.$

\item The limit $\displaystyle\lim_{r\to 0}\frac{A}{\rho}$ exists.
\end{enumerate}

\end{description}

The theorems in \cite{Manolo} can be seen to be subcases of the ones introduced here.

\section{Proof of the theorems}

In order to achieve prolongability of causal geodesics we just have to impose that $\dot t$
remains finite for finite values of the affine parameter. The radial velocity, $\dot r$, need
not be considered since it cannot become singular if $\dot t$ is not singular too. The other
derivatives, $\dot z$ and $\dot \phi$, are quadratures of smooth functions of $t$ and $r$ and
therefore they only may turn singular if $t$ or $r$ become so. We shall focus on
future-pointing geodesics. The analysis for past-pointing geodesics is
entirely similar.

A way of preventing the hyperbolic functions of $\xi$ from becoming singular is to require that
 $\xi$ does not grow indefinitely. Therefore for large values of $\xi$, $\dot
\xi$ must eventually become negative.  Since the constants, $L,P,\delta$ may
vanish independently one is not to expect compensations between them. Hence
their respective terms in (\ref{xi}) have to become negative for large values of
positive $\xi$ as it is stated in conditions (\ref{Mxi1}) and (\ref{Mxi2}),
taking into account that $\cosh \xi=\sinh
\xi+e^{-\xi}$ and that therefore the terms in the negative exponential of $\xi$ (\ref{Mxi2})
need not be negative but just of the same order as those in (\ref{Mxi1}).

By imposing condition (\ref{ax}) in the theorem, we require that the geometry of the spacetime
in the vicinity of the axis is determined by $\rho$ and not by $A$. Hence for negative
decreasing
$\xi$ the axis could be singular only for geodesics with $L\neq 0$, since the terms $A/\rho$
are finite at the axis. Geodesics with zero angular momentum just cross $r=0$ and reappear
with positive
$\xi$ and polar angle $\phi+\pi$ and hence need not be taken into account. In this case we can
therefore allow compensations between the non-zero $L$ term and the other ones. Splitting
$\cosh\xi$ as $e^{\xi}-\sinh\xi$, the condition for $\dot \xi$ to become positive for large
values of $t$ and negative $\xi$ is stated in (\ref{mxi1}) for the terms in $\sinh\xi$ and in
(\ref{mxi2}) for the terms in $e^{\xi}$, that can be at most of the same order of the former
ones since they are exponentially damped. 

No further conditions need be imposed on $\dot\xi$. But $\dot t$ could turn singular also for
the $e^{-2g}F$ term. A way of preventing it is to impose a growth slower than linear for $\dot
t$ due to each of the three terms ($\delta$, $\Lambda$, $P$) for large values of $t$. This is
done in condition (\ref{tt}).

The condition on $g$ must be refined since we have not yet considered the geodesics that
cannot be parametrized by $\xi$. These are those with zero $F$, that is, null geodesics with
zero $\dot z$ and $\dot \phi$. Since $\dot t=|\dot r|$ for such geodesics, the equations of
geodesic motion,
\begin{eqnarray}
\ddot t+g_t\dot t^{2}
+2\,g_r{\dot t}{\dot r}+g_t{\dot r}^{2}=0,
\end{eqnarray}
\begin{eqnarray}
\ddot r+g_r\dot t^{2}+2\,g_t{\dot t}{\dot r}+g_r{\dot r}^{2}=0,
\end{eqnarray}  
can be reduced to a single one that can be integrated,
\begin{eqnarray}
\ddot t+2\,g_t\dot t^{2}+2\,g_r{\dot t}{\dot r}=0\Rightarrow (e^{2g}\dot t)^\cdot=K,
\end{eqnarray}
which, in order to have $t$ extendible to arbitrary values of the affine parameter, can be
controlled by imposing at most linear growth for $\dot t$ as it is done in condition 
(\ref{tt}).

\section{Completeness of several cylindrical models}

Since all known diagonal cylindrical perfect fluid models
(\cite{Seno},\cite{Ruiz},\cite{Sep},\cite{leo}) with regular curvature invariants have already
been shown to be geodesically complete \cite{Manolo}, we shall only concern about nondiagonal
ones. To our knowledge there are just two and both can be derived from Einstein
spacetimes using the Wainwright-Ince-Marshman generation algorithm for stiff
perfect fluids \cite{Wain}. 

\begin{enumerate}
\item Mars: \cite{Diag} It is the first known nonsingular
nondiagonal cylindrical cosmological model in the literature. In another
context it was previously published by Letelier \cite{Letelier}. In isotropic
coordinates the metric functions can be written as,

\begin{eqnarray}
&g(t,r)=\frac{1}{2}\ln\cosh(2\,a\,t)+\frac{1}{2}\alpha\,
a^2r^2,\qquad f(t,r)=\frac{1}{2}\ln\cosh(2\,a\,t),\nonumber\\ 
&\rho(t,r)=r,\qquad A(t,r)=a\,r^2,
\end{eqnarray}
where $a$ is a constant and $\alpha>1$. If $\alpha=1$ the pressure and the
density of the fluid vanish.

All functions are even in $t$ and therefore the analysis of past-pointing
geodesics is equivalent to that of future-pointing ones and shall be omitted.

The derivatives $g_u$, $q_u$ in condition (\ref{Mxi1}) are positive for positive
$t$ whereas $h_u$ also needs large radial coordinate. 

The derivatives $g_r$, $q_r$ in condition (\ref{Mxi2}) are positive regardless of 
$t$ and $h_r$ is positive for large $r$. 

Conditions (\ref{mxi1}), (\ref{mxi2})  are satisfied  when
 $r$ is small and $t$ is positive since the $L$ term is dominant for small $r$.

The functions in  condition (\ref{tt}) are all positive except for the 
$\ln r$ term in $h$ when $r$ decreases. But this can be bounded by a logarithm of $t$ and
therefore the condition is fulfilled.

The ratio $A/\rho$ tends to zero for decreasing $r$ and hence condition 
(\ref{ax}) in theorem $1$ is satisfied.

Hence this spacetime is causally $g$-complete.

\item Griffiths-Bi\v{c}\'ak: \cite{Jerry} The previous model is comprised in
this one for $c=0$ after a redefinition of constants. The metric functions can
be written as,

\begin{eqnarray}
&g(t,r)=\frac{1}{2}\ln\cosh(2\,a\,t)+\frac{1}{2}
a^2r^2+\frac{1}{2}\Omega(t,r),\ 
f(t,r)=\frac{1}{2}\ln\cosh(2\,a\,t),\nonumber\\  &\rho(t,r)=r,\qquad
A(t,r)=a\,r^2,
\end{eqnarray}
where $\Omega$ is a function that is obtained from a solution, $\sigma$, of the
wave equation,
\begin{eqnarray}
&\Omega_r=r(\sigma_t^2+\sigma_r^2),\qquad \Omega_t=2r\,\sigma_t
\sigma_r,\nonumber\\
&\sigma(t,r)=bt+\sqrt{2}{c}\sqrt{\frac{\sqrt{(\alpha^2+r^2-t^2)^2
+4\alpha^2t^2}+\alpha^2+r^2-t^2}{(\alpha^2+r^2-t^2)^2
+4\alpha^2t^2}}.
\end{eqnarray}

The analysis of the geodesics of this spacetime can be dealt with easily
since,
\begin{equation}
\Omega_u=r(\sigma_t+\sigma_r)^2,\qquad \Omega_v=-r(\sigma_t-\sigma_r)^2,
\end{equation}
and therefore $\Omega$ contributes with an additional positive term to conditions
(\ref{Mxi1}), (\ref{Mxi2}) in theorem 1, that were already checked for Mars
spacetime.

On the contrary, $\Omega$ adds a negative term in conditions (\ref{mxi1}),
(\ref{mxi2}) but it is negligible for small $r$.

Finally, the contribution of $\Omega$ to $g$ grows at most as $\sqrt{t}$ for
large $t$ and is therefore negligible compared to the other terms in condition
(\ref{tt}).

A similar reasoning is valid for theorem 2, although the metric function
$\Omega$ is not even in time.

Hence these spacetimes are geodesically complete.

\end{enumerate}

\vspace{0.3cm}
\section*{Acknowledgements}
The present work has been supported by Direcci\'on
General de Ense\~nanza Superior Project PB95-0371. The author wish to thank
 F. J. Chinea. L. M. Gonz\'alez-Romero and M. J. Pareja for valuable discussions
.

\end{document}